# A $^{57}$Fe Mössbauer Spectroscopy Study of the 7 K Superconductor LaFePO

Marcus Tegel[a], Inga Schellenberg[b], Rainer Pöttgen[b], and Dirk Johrendt[a,] *

[a] Department Chemie und Biochemie, Ludwig-Maximilians-Universität München, Butenandtstrasse 5–13 (Haus D), D-81377 München, Germany

[b] Institut für Anorganische und Analytische Chemie, Universität Münster, Corrensstrasse 30, D-48149 Münster, Germany

Reprint requests to D. Johrendt. E-mail: dirk.johrendt@cup.uni-muenchen.de

((**Heading:** M. Tegel *et al.* · $^{57}$Fe Mössbauer Spectroscopy on LaFePO))

A polycrystalline sample of superconducting LaFePO was prepared in a tin flux at 1123 K. The structure was determined from single crystal data (ZrCuSiAs-type, *P*4/*nmm*, *a* = 3.9610(1), *c* = 8.5158(2) Å, *Z* = 2) and the phase analysis was performed by the Rietveld method. LaFePO is Pauli-paramagnetic and becomes superconducting at 7 K after removing the ferromagnetic impurity phase $Fe_2P$ from the sample. $^{57}$Fe Mössbauer spectroscopy measurements at 298, 77, 4.2 and 4 K show single signals at isomer shifts around 0.35 mm/s, subject to weak quadrupole splitting. At 4 K, a symmetric line broadening appears, resulting from a small transferred magnetic hyperfine field of 1.15(1) T and accompanied by an angle of 54.7(5)° between $B_{hf}$ and $V_{zz}$, the main component of the electric field gradient tensor.

*Key words*: Pnictide Oxide, Iron, Superconductivity, Mössbauer Spectroscopy

**Introduction**

The recent discovery of high-$T_C$ superconductivity in the fluoride doped quaternary iron arsenide oxides $RE$FeAs(O$_{1-x}$F$_x$) ($RE$ = La-Sm) [1] with critical temperatures ($T_C$) up to 55 K [2] has sparked a tremendous interest in compounds with the ZrCuSiAs-type structure [3]. The parent compound LaFeAsO does not become superconducting, but shows a spin density wave (SDW) anomaly at 150 K and antiferromagnetic ordering below 138 K [4, 5]. When doping the oxide site with fluoride, the SDW is suppressed and superconductivity occurs at critical temperatures up to 41 K in the iron arsenide oxide LaFeAs(O$_{1-x}$F$_x$) [6]. On the other hand, the iron phosphide oxide LaFePO had previously been reported to become superconducting at 3.5-4.1 K even in the undoped case [7, 8]. Upon fluoride doping, the $T_C$ had been reported to increase slightly to 6 K. Such large differences in $T_C$ between the phosphide and arsenide oxides do not occur in the isotypic nickel compounds LaNiPO [9,10] and LaNiAsO [11]. Both compounds are superconductors around 2-4 K and up to now, their low transition temperatures could not be increased significantly by doping.

These results emphasize the exceptional position of the iron arsenide oxides. One key factor for their higher $T_C$ obviously is the existence of a SDW in the undoped phase, which becomes unstable around 150 K, before it locks into an antiferromagnetic spin ordering. This suggests that spin fluctuations play an important role in the pairing mechanism, similar as in the high-$T_C$ cuprates. It is not known up to now, if a comparable magnetic anomaly also occurs in superconducting LaFePO. However, the much lower $T_C$ and its insignificant increase upon doping could imply that the pairing mechanism is different in phosphide and arsenide oxides.

Recent investigations on the magnetic properties of doped and undoped LaFeAsO by $^{57}$Fe Mössbauer spectroscopy have proved spin ordering in LaFeAsO and its suppression upon doping [12, 13]. In order to shed light on a potentially different nature of superconductivity in the corresponding phosphide oxide, we present $^{57}$Fe Mössbauer spectra, magnetic measurements and structural details of LaFePO in this paper.

**Experimental**

*Synthesis*

A polycrystalline sample of LaFePO was synthesized by heating a mixture of La, Fe, $Fe_2O_3$ and red P in a ratio of 3:1:1:3 (a total of ~650 mg) in a tin flux (2.5 g tin) [14] under argon atmosphere by using an alumina crucible in a silica ampoule. The sample was heated at 1123 K (40 K/h) for 7 days and slowly cooled to room temperature (15 K/h). The tin ingot was dissolved in 6 M HCl at room temperature. This procedure yielded a black powder consisting of platelet crystals with a metallic lustre.

*X-Ray diffraction*

X-ray powder patterns of the LaFePO sample were recorded on a Stoe Stadi-P diffractometer (Mo-$K_{\alpha 1}$, Ge(111)-monochromator, $\lambda$ = 70.93 pm, Si as external standard). The powder data were analyzed by the Rietveld method using the GSAS suite [15]. Single crystal data of LaFePO were collected using an Enraf-Nonius κ-CCD equipped with a rotating anode (Mo-$K_\alpha$ radiation, $\lambda$ = 71.073 pm). Intensities were corrected for absorption with SADABS [16] and refined against $F^2$ using SHELXL97 [17] with anisotropic displacement parameters for all atoms except oxygen. Starting parameters were taken from isostructural PrFePO [18]. Details about the crystal structure determination may be obtained from: Fachinformationszentrum Karlsruhe, D-76344 Eggenstein-Leopoldshafen, Germany, e-mail: crysdata@fiz-karlsruhe.de, on quoting the registration No. CSD-391428.

*Magnetic measurements*

Magnetization measurements were performed using a Quantum-Design SQUID magnetometer (MPMS-XL5) between 1.8 and 300 K. A powdered sample of LaFePO was placed into a gelatin capsule and fixed in a straw as sample holder. Zero-field-cooling (Shielding) and field-cooling (Meissner) measurement cycles were performed at 10 Oe between 1.8 and 12 K in the reciprocating sample option (RSO) mode. The magnetic susceptibility between 2 and 300 K was measured at 10 kOe.

*$^{57}$Fe Mössbauer spectroscopy*

A $^{57}$Co/Rh source was available for the $^{57}$Fe Mössbauer spectroscopy investigations. The LaFePO sample was placed in a thin-walled PVC container at a thickness of about 10 mg Fe/cm$^2$. The measurements were run in the usual transmission geometry at 298, 77, 4.2, and 4 K. The source was kept at room temperature.

**Results and Discussion**

The ZrCuSiAs-type structure [19] was confirmed by single crystal data. LaFePO is build up by LaO- and FeP- layers alternating along [001]. Lanthanum is eightfold coordinated by four oxygen (2.355 Å) and four phosphorous atoms (3.343 Å). The Fe–P bonds lengths (2.295 Å) are close to the sum of the covalent radii of iron and phosphorous (2.26 Å, [20]). The FeP$_4$ tetrahedra are slightly flattened along [001], as it can be recognized by a P–Fe–P angle of 119.3°. This distortion is slightly bigger than in LaFeAsO with an As–Fe–As angle of 113.7° (at 175 K) [4], but much weaker than that of the NiP$_4$ tetrahedra in LaNiPO, where the P–Ni–P angle is 126.5° [9].

The Rietveld analysis revealed that the sample was not single phase, small amounts of Fe$_2$P and FeSn$_2$ could be detected. Such magnetic impurity phases can be very destructive for susceptibility measurements. Fe$_2$P is ferromagnetic below 266 K [21] and FeSn$_2$ is antiferromagnetic below 380 K [22], thus even small traces of Fe$_2$P would strongly affect the magnetic measurements. This is especially important with respect to the weak Pauli-paramagnetic LaFePO. Indeed, our first susceptibility measurement of the LaFePO sample showed a strong upturn of $\chi$ below 270 K, which coincides with the Curie point of Fe$_2$P. Also no superconductivity could be detected at temperatures down to 1.8 K. In order to remove impurities, we separated ferromagnetic particles by stirring a suspension of the finely grounded LaFePO sample in liquid N$_2$ with a strong permanent magnet. After this treatment, the Fe$_2$P impurity was drastically reduced and we could successfully fit the complete X-ray powder pattern with phase fractions of 96 % LaFePO and 4 % FeSn$_2$, respectively. However, the magnetic measurement of this purified sample still revealed small traces of a residual Fe$_2$P impurity, as depicted in Figure 1. But now the sample exhibited the typical strong diamagnetic shielding of a superconductor. The complete shielding- and Meissner-cycle (Figure 2) clearly shows the onset of superconductivity at 7 K, which is significantly higher than 4.1 K as it was reported for undoped LaFePO [7, 8]. Obviously, the superconductivity has been completely suppressed by the ferromagnetic impurity phase, which generates a magnetic field inside the sample. If the critical field of the superconducting phase is sufficiently small, this additional field can decrease $T_C$ or even bring it to zero. Our magnetization measurements at 1.8 K showed small critical fields of $H_{c1} \approx 75$ Oe and $H_{c2} \approx 880$ Oe for LaFePO, which correspond approximately to the values given by *Hosono* [8].

Our results indicate that ferromagnetic impurities can strongly influence the superconductivity in LaFePO and presumably also in the LaFeAsO compounds. This may be one reason for the partially different $T_C$'s of supposedly identical compounds, which are in most cases far from being single phase. However, in the case of fluoride doped compounds, the exact amount of fluoride in the structure may be another important problem.

$^{57}$Fe Mössbauer spectra of LaFePO recorded at 298, 77, 4.2, and 4 K are presented in Figure 3 together with transmission integral fits. The corresponding fitting parameters are listed in Table 2. In agreement with the ZrCuSiAs-type crystal structure, the spectra were well reproduced with single iron sites at isomer shifts around 0.3 mm/s, slightly smaller than the isomer shifts observed recently for LaFeAsO and LaFeAsO$_{0.89}$F$_{0.11}$ [12,13]. Due to the non-cubic site symmetry, the spectra are subject to weak quadrupole splitting.

Similar to LaFeAsO and LaFeAsO$_{0.89}$F$_{0.11}$, also for LaFePO we observe a slight increase of the isomer shift with decreasing temperature. For iron, a smaller isomer shift is consistent with a higher electron density at the nuclei [23]. The isomer shifts observed for LaFePO are comparable with other iron phosphides with tetrahedrally coordinated iron [24, 25].

At 298 and 77 K, we observe no magnetic hyperfine field splitting, clearly manifesting the absence of magnetic ordering, similar to the fluoride doped superconducting arsenide LaFeAsO$_{0.89}$F$_{0.11}$ [12, 13]. Since a strong anomaly has been observed in the specific resistivity of LaFeAsO [5], associated with a spin density wave, and full magnetic hyperfine field splitting below this phase transition is evident from $^{57}$Fe Mössbauer spectroscopy, we suppose that the mechanism for superconductivity in LaFeAsO and LaFePO is different. This is also supported by the facts, that the parent compound LaFeAsO is not superconducting while LaFePO is, and that LaFeAsO becomes superconducting at high $T_C$ upon doping, whereas $T_C$ increases only insignificantly in doped LaFePO. Different mechanisms of superconductivity for iron and nickel pnictide oxides have also been suggested by theoretical arguments [26].

The 4 K spectrum shows a symmetric line broadening. A reliable fit was obtained by simultaneously applying a weak quadrupole splitting of 0.15(1) mm/s and a transferred magnetic hyperfine field ($B_{hf}$) of 1.15(1) T. In order to explain the symmetric spectrum with a combined hyperfine field and an electrical quadrupole interaction, the angle $\theta$ between $B_{hf}$ and $V_{zz}$ (the main component of the electric field gradient tensor) should be close to the magic angle. Indeed, the refined $\theta$ value for LaFePO was 54.7(5)°. This behavior has also been observed for other magnetically ordered rare earth phases [27, 28].

A small hint of this behavior was already evident in the 4.2 K spectrum, however, the very small transferred hyperfine field is hidden in the slightly increased line width and quadrupole splitting parameters, hampering independent refinement of the hyperfine field parameter. Although our experimental setup is limited to 4 K, we expect higher transferred hyperfine fields at lower temperatures.

*Acknowledgements*

This work was financially supported by the Deutsche Forschungsgemeinschaft. We thank Dipl.-Chem. F. M. Schappacher for help with the Mössbauer spectroscopy.

Table 1. Crystal structure data for LaFePO.

| Formula | LaFePO | $\mu(MoK_\alpha)$, cm$^{-1}$ | 21.454 |
|---|---|---|---|
| $M_r$ | 241.726 | $hkl$ range | $h,k \pm 5, l \pm 11$ |
| Cryst. size, mm$^3$ | 0.043 x 0.05 x 0.015 | $(\sin\theta/\lambda)_{max}$, Å$^{-1}$ | 1.152 |
| Crystal system | tetragonal | Refl. measured | 1976 |
| Space group | $P4/nmm$ | Refl. unique | 102 |
| $a$, Å | 3.9610(1) | $R_{int}$ | 0.107 |
| $c$, Å | 8.5158(2) | Param. refined | 11 |
| $V$, Å$^3$ | 133.61(1) | $R(F)/wR(F^2)^a$ (all) | 0.043, 0.105 |
| $Z$ | 2 | GoF $(F^2)^a$ | 1.173 |
| $D_{calcd}$, g cm$^{-3}$ | 6.07 | | |

Atomic positions and displacement parameters

| Atom | Wyck. | $z$ | $U_{eq}$ (pm$^2$) |
|---|---|---|---|
| La | 2$c$ (¼, ¼, $z$) | 0.1496(2) | 96(8) |
| Fe | 2$b$ (¾, ¼, ½) | ½ | 91(9) |
| P | 2$c$ (¼, ¼, $z$) | 0.6362(6) | 93(13) |
| O | 2$c$ (¾, ¼, 0) | 0 | 60(30) |

Selected bond lengths (Å) and angles (deg)

| La–O (4×) | 2.355(1) | Fe–P (4×) | 2.295(3) |
|---|---|---|---|
| La–P (4×) | 3.343(3) | Fe–Fe (4×) | 2.801(1) |
| La–O–La | 107.0(1), 114.5(1) | P–Fe–P | 104.8(1), 119.3(2) |

[a] Definition of $R$ values and GoF, as well as information on weighting scheme applied

Table with Mössbauer fitting parameters.

Table 2. Fitting parameters for $^{57}$Fe Mössbauer spectroscopy measurements of LaFePO. $\delta$: isomer shift; $\Gamma$: experimental line width; $\Delta E_Q$: quadrupole splitting parameter.

| $T$ (K) | $\delta$ (mm / s) | $\Gamma$ (mm / s) | $\Delta E_Q$ (mm / s) |
|---|---|---|---|
| 298 | 0.24(1) | 0.32(3) | 0.11(3) |
| 77  | 0.34(1) | 0.28(4) | 0.12(3) |
| 4.2 | 0.36(1) | 0.37(3) | 0.19(2) |
| 4   | 0.36(1) | 0.32(1) | 0.15(1) |

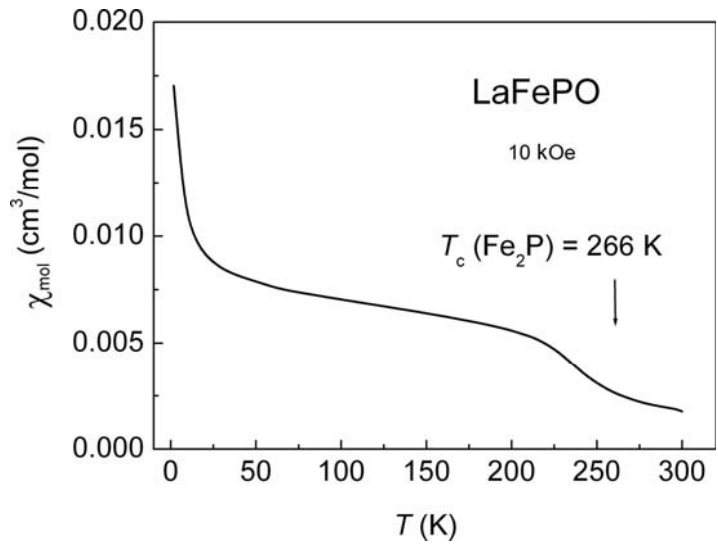

Figure 1. Magnetic susceptibility of the LaFePO sample measured at 10 kOe.

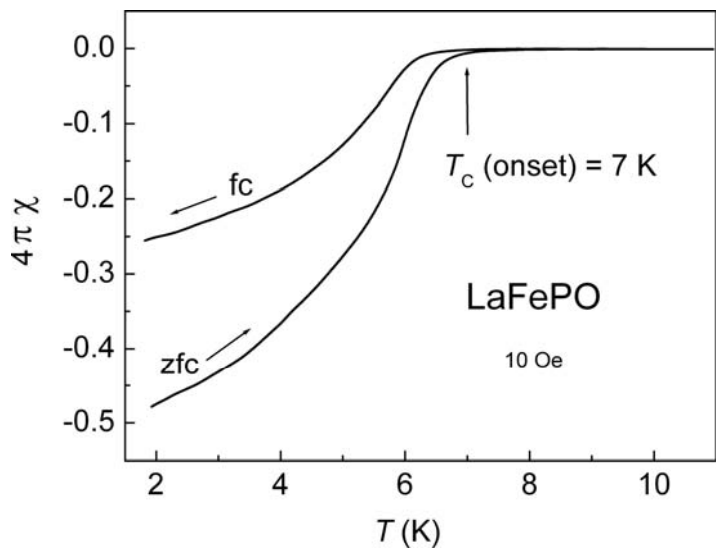

Figure 2. Diamagnetic shielding (zfc) and Meissner-Effect (fc) of LaFePO.

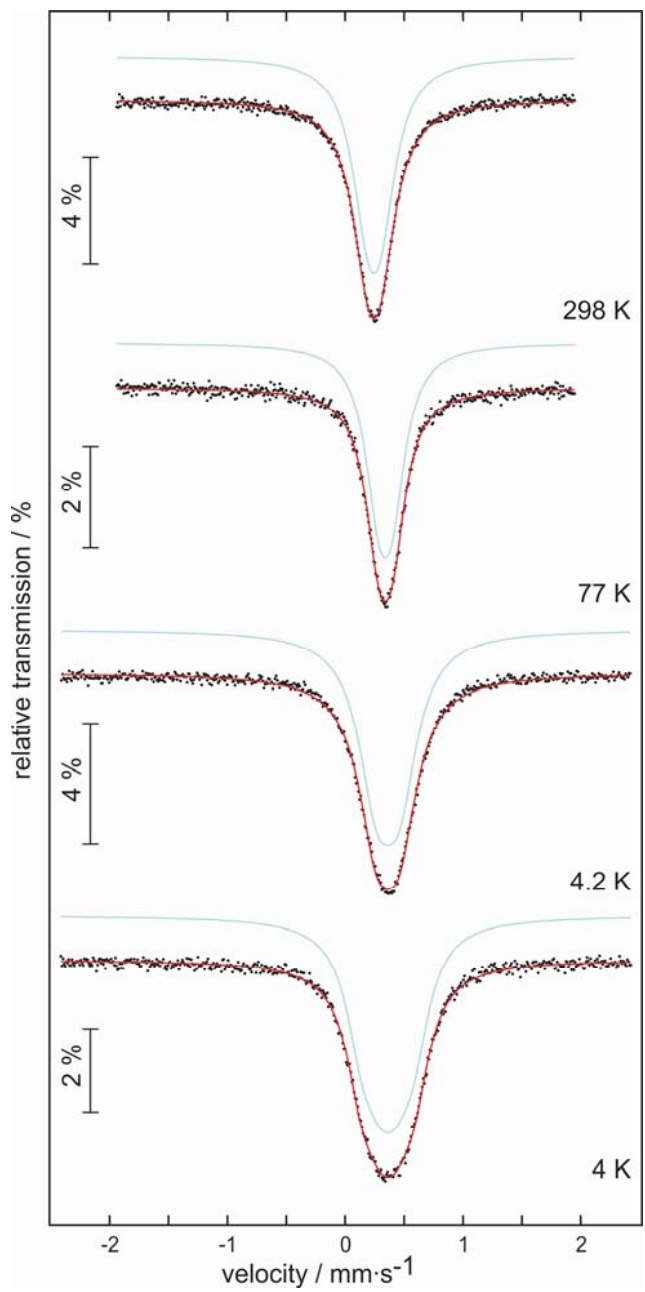

Figure 3. Experimental and simulated [57]Fe Mössbauer spectra of LaFePO at 298, 77, 4.2 and 4 K.